\documentclass[12pt,showpacs,aps]{revtex4}

\begin{document}
\title{On the vacuum fluctuations and the cosmological constant: Comment on the paper by T.Padmanabhan}

\author{V.G.Gurzadyan$^{1,2}$ and S.-S. Xue$^{2}$}

\vspace{0.1in}
\email{gurzadya@icra.it; xue@icra.it}

\affiliation{(1) Yerevan Physics Institute, Armenia \\
(2) ICRA,
Physics Department, University of Rome ``La Sapienza", Rome, Italy}


\maketitle

{\bf Abstract}

The formula for the dark energy, derived in recent Letter by Padmanabhan \cite{P} (given in its Abstract) was actually derived 4 years earlier in our paper \cite{GX}. Among dozens of references in \cite{P}, no quotation to our paper. 
Based on the same Zeldovich idea on vacuum fluctuations, Padmanabhan derives it from scaling considerations, while we had gone into more details and shown that the formula fits the observed value of the cosmological constant if l=0 modes are relevant.  

\vspace{0.2in}

Padmanabhan  derives (Eq.(6) in \cite{P})
\begin{equation}
\delta \rho_{vac} \propto L_p^{-2} L_H^{-2},
\end{equation}
where $L_p$ is the Planck lenght and $L_H$ is the Hubble scale.

In \cite{GX} we had shown (Eqs. (17-23) in the second paper in \cite{GX}), that if the condition $l=0$ for the angular quantum number is adopted for relevant vacuum modes in the cosmological constant expression
\begin{equation}
\Lambda = 8\pi G \hbar c\sum_l{(2l+1)\over a^2(t)}\int{dk_r\over 2(2\pi)}
\sqrt{k_r^2+{l(l+1)\over a^2}},
\end{equation}
then one gets for the dark energy density 
\begin{equation}
\rho_\Lambda\equiv {\Lambda\over 8\pi G}= {1\over4}{\hbar c\pi\over a^4}
N_{\max}(N_{\max}+1),
\end{equation}
where $N_{\max}$ is the maximum number of relevant
radial modes
\begin{equation}
N_{\max}\simeq {a\over L_{p}}\simeq 10^{61},
\end{equation}
and the present characteristic size of the Universe $a\simeq 1\cdot 10^{28}\, cm$. The condition $l=0$ can be interpreted
as micro appearance of FRW metric. Such vacuum mode/cosmological constant link is the original
idea of Zeldovich \cite{z}.

This yields (Eq.(22) in \cite{GX}) for the dark energy density
\begin{equation}
\rho_\Lambda =  \frac{\hbar c \pi}{8a^2 L_{p}^2}=\frac{\pi c^4}{8a^2 G}
\end{equation}
and numerically as an effective matter density
\begin{equation}
\rho_{m\Lambda}=\frac{\pi c^2}{8a^2 G} \sim
10^{-29}\, g \, {\rm cm}^{-3}.
\end{equation}
The corresponding vacuum pressure is negative and thus accelerates
the expansion of the Universe. Note, that $\hbar$ is absent in Eqs.(5,6),
and also $G$ is absent in formula for $\Lambda$.

The remarkable numerical coincidence of the obtained density with that
of indicated by the supernovae observations, points out on the 
the fine tuning between the macro and micro structure of the Universe
which has to follow from future more fundamental theories.  
Some consequences of Eqs.(5,6) are in \cite{v}.

\end{document}